\documentclass[12pt]{article}
\usepackage[english]{babel}
\usepackage{bm}
\usepackage{amssymb}
\usepackage{amsmath}
\usepackage{amsfonts}
\usepackage{dsfont}
\usepackage{amsbsy}
\usepackage{graphicx}

\newcommand{\slsh}[1]{\not{\hbox{\kern-2pt${#1}$}}}

\newcommand{\ba}[1]{\begin{eqnarray} \label{#1}}
\newcommand{\ea}{\end{eqnarray}}

\def\bea{\begin{eqnarray}}
\def\eea{\end{eqnarray}}
\def\bqu{\begin{quote}}
\def\equ{\end{quote}}
\newcommand{\newc}{\newcommand}
\newc{\ra}{\rightarrow}
\newc{\lra}{\leftrightarrow}

\def\la{\lambda}

\newc{\sm}{Standard Model}
\newc{\smd}{Standard Model}
\newc{\barr}{\begin{eqnarray}}
 \newc{\earr}{\end{eqnarray}}

\def\gappeq{\mathrel{\rlap {\raise.5ex\hbox{$>$}}
{\lower.5ex\hbox{$\sim$}}}}

\def\lappeq{\mathrel{\rlap{\raise.5ex\hbox{$<$}}
{\lower.5ex\hbox{$\sim$}}}}

\begin{document}
\pagestyle{empty}

\begin{flushright}
KCL-PH-TH/2013-02, LCTS/2013-01, CERN-PH-TH/2013-02 \\
UHU-GEM/2013-07\\

\end{flushright}

\vspace*{1 cm}
\begin{center}
{\Large {\bf 
Neutrino textures and charged lepton flavour violation \\
\vspace*{0.2cm}
in light of $\theta_{13}$, MEG and LHC data}} \\
\vspace*{1cm}
{\bf M. Cannoni$^1$, J.~Ellis$^{2,3}$, M.E. G{\'o}mez$^1$} and {\bf S.~Lola$^{\star}$
\let\thefootnote\relax\footnotetext{$^{\star}$On leave from the Department of Physics, 
University of Patras, 26500 Patras, Greece.}} \\
\vspace{0.3cm}
$^1$ Department of Applied Physics, University of Huelva, 21071 Huelva, Spain \\
$^2$ Theoretical Particle Physics and Cosmology Group, Department of Physics, \\
King's College London, Strand, London WC2R 2LS, UK \\
$^3$ Theory Division, Physics Department, CERN, CH-1211 Geneva 23, Switzerland
\vspace*{0.3cm}
%
%
%
%
%
%
\vspace*{1.2cm}
{\bf ABSTRACT} \\ 
\end{center}
In light of recent results from the LHC, MEG and  neutrino experiments, we 
revisit the issue of charged lepton flavour violation (LFV)
in supersymmetric theories with massive neutrinos, where flavour-violating soft 
supersymmetry-breaking masses for sleptons are induced naturally by radiative corrections. 
We link our results to the expectations for light neutrinos with a normal mass 
hierarchy in SU(5), enhanced by an abelian flavour symmetry,
with particular focus on $\theta_{13}$.
We focus on the radiative 
decays $\ell_i \rightarrow \ell_j \gamma$ 
and on detection prospects at the LHC and a linear collider (LC). 
We use supersymmetric parameters consistent with 
cosmological considerations and with LHC searches for supersymmetry
and the Higgs mass.
We find a class of scenarios where the LHC may be sensitive to LFV sparticle decays and LFV
processes could be detectable at a LC with centre-of-mass energy above 1~TeV, whereas
LFV lepton decays may be suppressed by cancellations in the decay
amplitudes. 

\vspace*{5cm}
\noindent

\setcounter{page}{1}
\pagestyle{plain}

\section{Introduction}

In recent years, the existence of neutrino masses and
oscillations with near-maximal  $\nu_\mu - \nu_\tau$
and large $\nu_e \to \nu_{\mu}$ mixing has been established by
extensive input from atmospheric~\cite{skatm}, solar~\cite{sksol} and
long-baseline reactor~\cite{KamLand} and accelerator~\cite{K2K,MINOS}
neutrino experiments. Initial input on the possible range of the $\theta_{13}$~\cite{fogli,GonzalezGarcia}
was provided by the T2K~\cite{T2K2011} and MINOS collaborations~\cite{MINOS2011}, and 
definitive evidence for a non-zero value of $\theta_{13}$ has been provided by
the reactor experiments Daya Bay~\cite{daya-bay} and RENO~\cite{reno}, and very recently
also by Double Chooz~\cite{double chooz}.

A natural expectation in theories with massive neutrinos is charged-lepton-flavour violation (LFV), 
which is enhanced in  supersymmetric theories via the renormalization of soft
supersymmetry-breaking parameters. The link between neutrino oscillations and violations of the
individual lepton numbers $L_{e, \mu, \tau}$ raises the prospect of observing processes such as 
$\mu \rightarrow e \gamma$, 
$\mu \rightarrow 3 e$, 
$\tau \rightarrow \mu \gamma$ and $\mu \to e$
conversion on heavy nuclei~\cite{LFV-revs}.
The present experimental upper limits on the most
interesting of these processes, summarised below, already constrain
significantly the parameter spaces of theoretical models:
\begin{eqnarray}
BR(\mu \ra e \gamma)     &<& 5.6 \times 10^{-13}   \;\;   \cite{MEG2013},   \\
BR(\tau \ra \mu \gamma)  &<& 4.4 \times 10^{-8}  \;\; \cite{PDG},   \\
BR(\tau \ra e  \gamma)  &<& 3.3 \times 10^{-8}   \;\;    \cite{PDG}. 
\end{eqnarray}
The strongest constraint on radiative decays is the recent 
MEG upper limit on BR$(\mu \ra e \gamma)$~\cite{MEG2013},
four times more stringent limit than the previous one~\cite{MEG2011}.

Within the supersymmetric framework, one should also keep in mind other possibilities for
observing LFV processes, such as
slepton pair production at a Linear Collider (LC)~\cite{sleptonscemu,LFV-LCa,LFV-LCb,LC2,
LFV-LC2,CannoniLC,CEGL-LC,Abada} and signals 
at the LHC~\cite{LHC1,LHCh, LHC2,Bartl,CEGLR2,AbadaLHC,EstevesLHC,HirschLHC}, particularly in 
$\chi_2\to \chi + e^\pm \mu^\mp$ 
$\chi_2\to \chi + \mu^\pm \tau^\mp$ decays 
(here $\chi$ is the lightest
neutralino, assumed to be the lightest supersymmetric particle (LSP),
and $\chi_2$ is the second-lightest neutralino).
These decays could provide search prospects
that are complementary to direct searches for flavour-violating decays
of charged leptons, particularly for heavy superparticle spectra.

In this paper we re-evaluate the
prospects for observable  charged LFV,
based on updated knowledge of neutrino mass and mixing parameters that
includes the recent measurement of $\theta_{13}$.
We work within the framework of the most natural
mechanism for obtaining hierarchical light neutrino masses, namely
the see-saw mechanism~\cite{seesaw}, in which
an effective Majorana mass matrix for light neutrinos,
$m_{eff}=m^D_{\nu}\cdot (M_{N})^{-1}\cdot m^{D^{ T}}_{\nu}$, arises
from Dirac neutrino masses $m_{\nu}^D$ of the same order as
the charged-lepton and quark masses, and heavy Majorana
masses $M_{N}$.
In supersymmetric theories, the neutrino Dirac couplings
$Y_\nu$ renormalise the soft supersymmetry-breaking sneutrino and slepton masses,
generating LFV in a natural way~\cite{bm}. Even if the soft
scalar masses were universal  at the unification scale, 
quantum corrections between the GUT scale and low energies
would modify this structure via renormalization-group 
running, which generates off-diagonal contributions.  This effect is particularly interesting
in see-saw models, where in general the Dirac neutrino Yukawa couplings
cannot be diagonalized simultaneously with the charged-lepton 
and slepton mass matrices~\cite{bm}.
Given the large mixing of the corresponding neutrino species,
charged LFV may occur at enhanced rates in 
supersymmetric extensions of the standard model,
giving rise to observable LFV signals~\cite{LFV-revs,LFVhisano,gllv,Mismatch,Antusch,LFVres}.

We analyse this possibility within the constrained minimal supersymmetric standard model (CMSSM)
with universal scalar, gaugino  masses and trilinear terms at the GUT scale
($m_0$, $M_{1/2}$ and $A_0$, respectively), using mass matrices that are
inspired by GUT models with abelian flavour symmetries~\cite{GGR2,EGL}.
These textures reproduce naturally the observed
fermion mass hierarchies and mixing angles
and may also have interesting implications for leptogenesis
  \cite{leptogenesis, leptogenesis-LFV, Giudice, DI}.
Despite their phenomenological appeal, however, there
are ambiguities and limitations due to the fact that the entries 
in the mass matrices are determined only up to $\mathcal{O}(1)$ numerical factors.

The paper is organized in the following way:
In Section~\ref{sec:2} we look at the theoretical and phenomenological predictions for
neutrino mass matrices.
In Section~\ref{sec:3.1} we discuss the origin of LFV in
representative
supersymmetric scenarios, in
Section~\ref{sec:3.2} we analyse 
the connection between the U(1) charges and LFV and in
Section~\ref{sec:3.3}  we discuss the numerical procedure and the
renormalization group runs.
The obtained mixing matrices are then used to study various LFV processes:
in Section~\ref{sec:3.4} we discuss radiative decays and the impact  
of the new $\mu \to e \gamma$ MEG bound on the parameter space of
interest;  
in Section~\ref{sec:3.5} we study LFV in $\chi_2$ decays at the
LHC, while LFV from slepton production and decay at a future LC is discussed in Section~\ref{sec:3.6}.
In Section~\ref{sec:4} we discuss possible implications for leptogenesis, and
finally in Section~\ref{sec:5} we summarize the main results of the paper.

\section{Neutrino Mass Textures inspired by SU(5)}
\label{sec:2}

Over the recent years, a plethora of textures have been proposed to explain the data on 
neutrino masses and mixing. The new data on
$\theta_{13}$ provide additional constraints, excluding certain possibilities and constraining others.
Rather than reviewing the vast literature on the subject, we choose a representative 
model that fits the fermion data and is well-motivated on theoretical grounds. Nevertheless, we 
try to keep the results as generic as possible, placing emphasis on the links between physical 
observables. We also keep in mind that several {\it a priori} different theoretical models may converge 
to similar phenomenology, since they are matched to the same data.

The example we choose is provided by a SU(5) GUT combined with 
family symmetries~\cite{GGR2,EGL}.
The mass matrices are constructed by looking at the field content
of the SU(5) representations, namely: three families of
$(Q,u^{c},e^{c})_{i} \in {\tt 10}$ representations, three families of
$(L,d^{c})_i  \in { \tt \overline{5}}$ representations, 
and heavy right-handed neutrinos in singlet representations. 
This model therefore has the following properties:
(i) the  up-quark mass matrix is symmetric, and 
(ii) the charged-lepton mass matrix is the transpose of the 
down-quark mass matrix, which relates the mixing 
of the left-handed leptons to that of the
right-handed down-type quarks. 
Since the observed Cabibbo-Kobayashi-Maskawa (CKM) mixing in the quark sector is due 
to a mismatch between the mixing of the left-handed up- and 
down-type quarks, it can be easily reconciled with a large atmospheric neutrino 
mixing.

Within this framework, and following for example~\cite{GGR2,EGL,pedro},  the 
Yukawa matrices have the form
\begin{equation}
{Y}_{u}\propto \left(
\begin{array}{ccc}
\varepsilon ^6 & \varepsilon ^5 & \varepsilon ^3 \\
\varepsilon ^5 & \varepsilon ^4 & \varepsilon ^2 \\
\varepsilon ^3 & \varepsilon ^2 & 1 \\
\end{array}
\right),\,\,\,
{Y}_{\ell}\propto 
{Y}_{d}^T\propto 
\left(
\begin{array}{ccc}
\varepsilon ^4 & \varepsilon ^3 & \varepsilon  \\
\varepsilon ^3 & \varepsilon ^2 & 1 \\
\varepsilon^3  & \varepsilon^2  & 1 \\
\end{array}
\right),\,\,\,
{Y}_{\nu}\propto \left(
\begin{array}{ccc}
\varepsilon ^{|1\pm n_1|} & \varepsilon ^{|1 \pm n_2|} 
& \varepsilon ^{|1 \pm n_3|} \\
\varepsilon ^{|n_1|} & \varepsilon ^{|n_2|} & \varepsilon ^{|n_3|} \\
\varepsilon ^{|n_1|} & \varepsilon ^{|n_2|} & \varepsilon ^{|n_3|} \\
\end{array}
\right) \, ,
\label{Yukawas}
\end{equation}
where $Y_{u,d,\ell,\nu}$ stand for the Yukawa couplings of 
quarks, charged leptons and neutrinos respectively, and
$n_i$ denote the U(1) charges of the heavy Majorana neutrinos.
The heavy Majorana mass matrix is then given by
\begin{equation}
M_{N}\propto \left(
\begin{array}{ccc}
\varepsilon ^{2|n_1|} & \varepsilon^{|n_1 +n_2|}  & \varepsilon^{|n_1 +n_3|}  \\
\varepsilon^{|n_1 +n_2|}  & \varepsilon^{2| n_2|}  & \varepsilon^{|n_2 +n_3|} \\
\varepsilon^{|n_1 + n_3|}  & \varepsilon^{|n_2 + n_3|} & \varepsilon^{2| n_3|} \\
\end{array}
\right) \, .
\label{mN}
\end{equation}
There is no unique choice for the right-handed neutrino charges
${n_1,n_2,n_3}$, and several possibilities may be compatible with the
low-energy neutrino data. We know, however, that the neutrino masses and mixing angles are 
related to the $\nu_L \nu_L$ contributions in the effective 
neutrino mass matrix 
\begin{equation}
m_{eff} \approx m^D_\nu \frac{1} {M_{N}} {m^D_\nu}^T \, ,
\label{eq:m_nu}
\end{equation}
which, if calculated from the matrices in (\ref{Yukawas}) and (\ref{mN}), is of the form:
\begin{equation}
m_{eff}\propto \left(
\begin{array}{ccc}
\varepsilon ^2 & \varepsilon  & \varepsilon  \\
\varepsilon  & 1  & 1 \\
\varepsilon  & 1  & 1 \\
\end{array}
\right) \, .
\label{meff}
\end{equation}

This form of $m_{eff}$  is quite natural in the simplest see-saw models with
a single expansion parameter and generic structures for the heavy and light Majorana 
mass matrices, due to cancellations that eliminate the dependences on the right-handed
charges. Its predictions have been extensively analysed from a phenomenological point
of view~\cite{guid,altarellireview,SatoYan,Hall,Vissani,Haba,Meloni,buchmuller,Merlo,Acosta}, 
and give a reasonable match to the data, provided there are no
cancellations of potentially large mixing
in the charged lepton sector. 
Among other predictions, $\theta_{13}$ turns out to be of the correct
order of magnitude. 
It is interesting to also note that, to lowest order in $\epsilon$,
$Y_\ell Y_\ell^\dagger$ has the same structure as $m_{eff}$,  namely
\begin{equation}
Y_{\ell} Y_{\ell}^\dagger \propto m_{eff}\propto \left(
\begin{array}{ccc}
\varepsilon ^2 & \varepsilon  & \varepsilon  \\
\varepsilon  & 1  & 1 \\
\varepsilon  & 1  & 1 \\
\end{array}
\right) \, .
\label{eq:meff1}
\end{equation}

The flavour mixing matrices are determined
by the following diagonalizations of the Dirac and Majorana mass matrices:
\begin{eqnarray}
{V_\ell}^T (Y_\ell {Y_\ell}^\dagger) V_\ell^* &=& \text{diag}(y_e^2,y_\mu^2,y_\tau^2),\\
{V_D}^T (Y_\nu {Y_\nu}^\dagger) V_D^* &=& \text{diag}(y^2_{\nu_1},y^2_{\nu_1},y^2_{\nu_3}),\\
{U_N}^T M_{N} U_N &=& \text{diag}(M_1,M_2,M_3),\\
{U_\nu}^T m_{eff} U_\nu &=& \text{diag}(m_{\nu_1},m_{\nu_2},m_{\nu_3}) \, .
\end{eqnarray}
In terms of the above matrices, the Maki-Nakagawa-Sakata (MNS) matrix is given by 
\begin{equation}
U_{MNS} \equiv U = V^\dagger_\ell U_\nu \, ,
\end{equation}
and can be parametrized as:
\begin{equation}
U= 
V\cdot \text{diag}(e^{-i\phi_1/2},e^{-i\phi_2/2},1) \, ,
\end{equation}
where
\begin{equation}
V=\left( 
\begin{array}{ccc}
c_{12}c_{13} & s_{12}c_{13} & s_{13}e^{-i\delta}\\
-c_{23}s_{12}-s_{23}s_{13}c_{12}e^{i\delta} &
c_{23}c_{12}-s_{23}s_{13}s_{12}e^{i\delta} &
s_{23}c_{13}\\
s_{23}s_{12}-c_{23}s_{13}c_{12}e^{i\delta} &
-s_{23}c_{12}-c_{23}s_{13}s_{12}e^{i\delta} &
c_{23}c_{13}
\end{array}
\right) \, ,
\end{equation}
and $c_{ij}$ and  $s_{ij}$ stand for $\cos\theta_{ij}$ and 
$\sin\theta_{ij}$, respectively.

\subsection{Predictions for Neutrino Observables}
\label{sec:2.1}

Within the above framework 
there are ambiguities in the choices of coefficients, limited
to a certain extent by requiring consistency with the 
experimental data. 
The match of the neutrino data to textures predicted by abelian flavour symmetries can be made 
by treating coefficients as 
random variables in Monte-Carlo scans of the multi-dimensional
parameter space,
in a statistical study of the probability that the textures can
naturally reproduced the measured 
angles and masses~\cite{SatoYan,Hall,Vissani,Haba,Meloni,buchmuller,Merlo}.

Here, being interested in matrices that are naturally consistent with the neutrino 
data, we proceed by taking the expansion parameter to 
be $\varepsilon=0.2$ and multiply the entries of $Y_\ell$, $Y_\nu$ and
$M_N$ in Eqs.~(\ref{Yukawas}) and (\ref{mN}) by coefficients  $\ell_{ij}$, $\nu_{ij}$ and $N_{ij}$ 
in the range $\pm [0.5, 2]$. In addition, we impose the following constraints:
\begin{enumerate}

\item[(i)] We select the charged lepton Yukawa coupling matrix $Y_\ell$ so that the correct 
charged-lepton mass hierarchies are reproduced, namely
\begin{equation}
\frac{m_\mu}{m_\tau} \sim 0.06  \; \; \;  \;\;
\frac{m_e}{m_\tau} \sim 2.5 \times 10^{-3} \, .
\label{eq:exp_lep}
\end{equation}
\item[(ii)]
$Y_\nu$ and $M_N$ are required to give 
a light neutrino mass matrix $m_{eff}$ of the form (\ref{meff}), with entries
that deviate by a factor $\in [0.5,2]$  from those in Eq. (\ref{meff}).

\item[(iii)] 
We impose normal hierarchy among the neutrino masses.
We fix $m_{\nu_3}\sim \sqrt{\Delta m^2_{\text{atm}}}\sim 0.05$ eV and require $0.16 <m_{\nu_2}/m_{\nu_3}<0.19$, 
$m_{\nu_1}<0.2 m_{\nu_2}$ consistent with the measured values of 
$\Delta m^2_{\text{sol}}$ and $\Delta m^2_{\text{atm}}$~\cite{neu-fits}.

\item[(iv)] We require the following range of mixing angles~\cite{GonzalezGarcia}:
\bea
0.27<&\sin^2\theta_{12}&<0.35, \nonumber \\
0.34<&\sin^2\theta_{23}&<0.67, \nonumber\\
0.018<&\sin^2\theta_{13}&<0.033 \, . 
\label{eq:exp_neu2}
\eea
The range on $\theta_{13}$ is consistent with the values reported by both~\cite{daya-bay, reno} at the
$3\sigma$ level. 

\item[(v)] We make a further selection by requiring that the hierarchy of 
eigenvalues of $Y_\ell Y_\ell^\dagger$ (which, as discussed above, has a
similar structure to $m_{eff}$  and $Y_\nu Y_\nu^\dagger$)
preserves the  order of the gauge  eigenstates. 
This reduces the density of solutions in the plots and implies large off-diagonal elements in 
both $V_\ell$ and $U_\nu$. 

\end{enumerate}

The selection of coefficients in the textures is performed so that
the above conditions are 
satisfied and the coefficients are chosen to be 
real in the range $\pm [0.5, 2]$. Given our ignorance of the CP-violating phase $\delta$, we focus on the case 
$\delta=0$ and do not include the Majorana phases $\phi_{1,2}$.

In Table~\ref{tab:table1} we provide two representative 
examples of our fits, which
will be used for our analysis below.  
We quote our predictions 
for  neutrino mixing angles with and without taking into account
renormalization effects. 
The RGE runs for the 
 ``see-saw'' MSSM are evaluated using the code {\it REAP} , described
 in Ref.~\cite{Antusch2} . 
The coefficients are taken at the GUT scale. 
We work with $tan\beta=45$, 
since this is the largest value that we will use in the numerical
computations of the next section
(and larger  $tan\beta=45$ results to larger corrections).
In all cases, the effect of varying 
$tan\beta$ in the range of values used in our examples (from 16 to 45) has
an impact of less than  2\% on the final value of the neutrino mixing angles.
 
\begin{table}[t!]
\resizebox{\textwidth}{!}{
\begin{tabular}{cccc}
\hline \hline
\\
Fit  & $Y_{\ell}$ & $Y_{\nu}$ & $M_{N}$\\
\hline
\\
1&$\left(\begin{array}{rrr}
\varepsilon^4 & -1.6\varepsilon^3 & 1.2\varepsilon\\
0.7\varepsilon^3 & 1.6\varepsilon^2 & -0.6\\
0.7\varepsilon ^3 & -1.7\varepsilon^2 & -1.3
\end{array}\right)$ &
$\left(
\begin{array}{rrr}
\varepsilon ^{|1\pm n_1|} & \varepsilon ^{|1 \pm n_2|} & -1\varepsilon ^{|1 \pm n_3|} \\
0.8\varepsilon ^{|n_1|} & \varepsilon ^{|n_2|} & -1.2\varepsilon ^{|n_3|} \\
-1.3\varepsilon ^{|n_1|} & \varepsilon ^{|n_2|} & 0.7\varepsilon ^{|n_3|} \\
\end{array}\right)$ &

$\left(
\begin{array}{ccc}
\varepsilon ^{2|n_1|} & \varepsilon^{|n_1 +n_2|}  & -1.7\varepsilon^{|n_1 +n_3|}  \\
\varepsilon^{|n_1 +n_2|}  & \varepsilon^{2| n_2|}  & \varepsilon^{|n_2 +n_3|} \\
-1.7\varepsilon^{|n_1 + n_3|}  & \varepsilon^{|n_2 + n_3|} & -\varepsilon^{2| n_3|} \\
\end{array}
\right)$ 
\\
\\
&$\sin^2\theta_{13}$=0.020(0.022), & $\sin^2\theta_{12}$=0.267 (0.274)  & $\sin^2\theta_{23}$=0.580 (0.613) \\
\hline 
\\
2&$\left(\begin{array}{rrr}
\varepsilon^4  & -1.5\varepsilon^3   & -2\varepsilon\\ 
\varepsilon^3  & -1.9\varepsilon^2  & 0.5\\
0.5\varepsilon^3  & -\varepsilon^2 & 0.75 
\end{array}\right)$&
$\left(
\begin{array}{ccc}
\varepsilon ^{|1\pm n_1|} & \varepsilon ^{|1 \pm n_2|} & -2\varepsilon ^{|1 \pm n_3|} \\
1.5\varepsilon ^{|n_1|} & \varepsilon ^{|n_2|} & -0.75\varepsilon ^{|n_3|} \\
1.9\varepsilon ^{|n_1|} & \varepsilon ^{|n_2|} & 1.5\varepsilon ^{|n_3|} \\
\end{array}\right)$&
%
%
$\left(
\begin{array}{ccc}
\varepsilon ^{2|n_1|} & \varepsilon^{|n_1 +n_2|}  & -1.9\varepsilon^{|n_1 +n_3|}  \\
\varepsilon^{|n_1 +n_2|}  &  \varepsilon^{2| n_2|}  &  \varepsilon^{|n_2 +n_3|} \\
-1.9\varepsilon^{|n_1 + n_3|}  & \varepsilon^{|n_2 + n_3|} & 1.9\varepsilon^{2| n_3|} \\
\end{array}
\right)$
\\
\\
&$\sin^2\theta_{13}$=0.017(0.022) & $\sin^2\theta_{12}$=0.278(0.310) & $\sin^2\theta_{23}$=0.390(0.439) \\
\hline \hline
\end{tabular} }
\caption{\it Indicative textures for $Y_{\ell}$, $Y_{\nu}$ and $M_N$
  at the GUT scale, to be studied in detail below. The $n_i$ are abelian
charges, that can only be constrained by LFV.
The computation of the neutrino 
mixing angles includes the RGE effects using $tan\beta = 45$ (in parenthesis we quote the 
predictions without the RGE runs). 
}
\label{tab:table1}
\end{table}

\begin{figure*}
\begin{center}
\includegraphics*[scale=0.43]{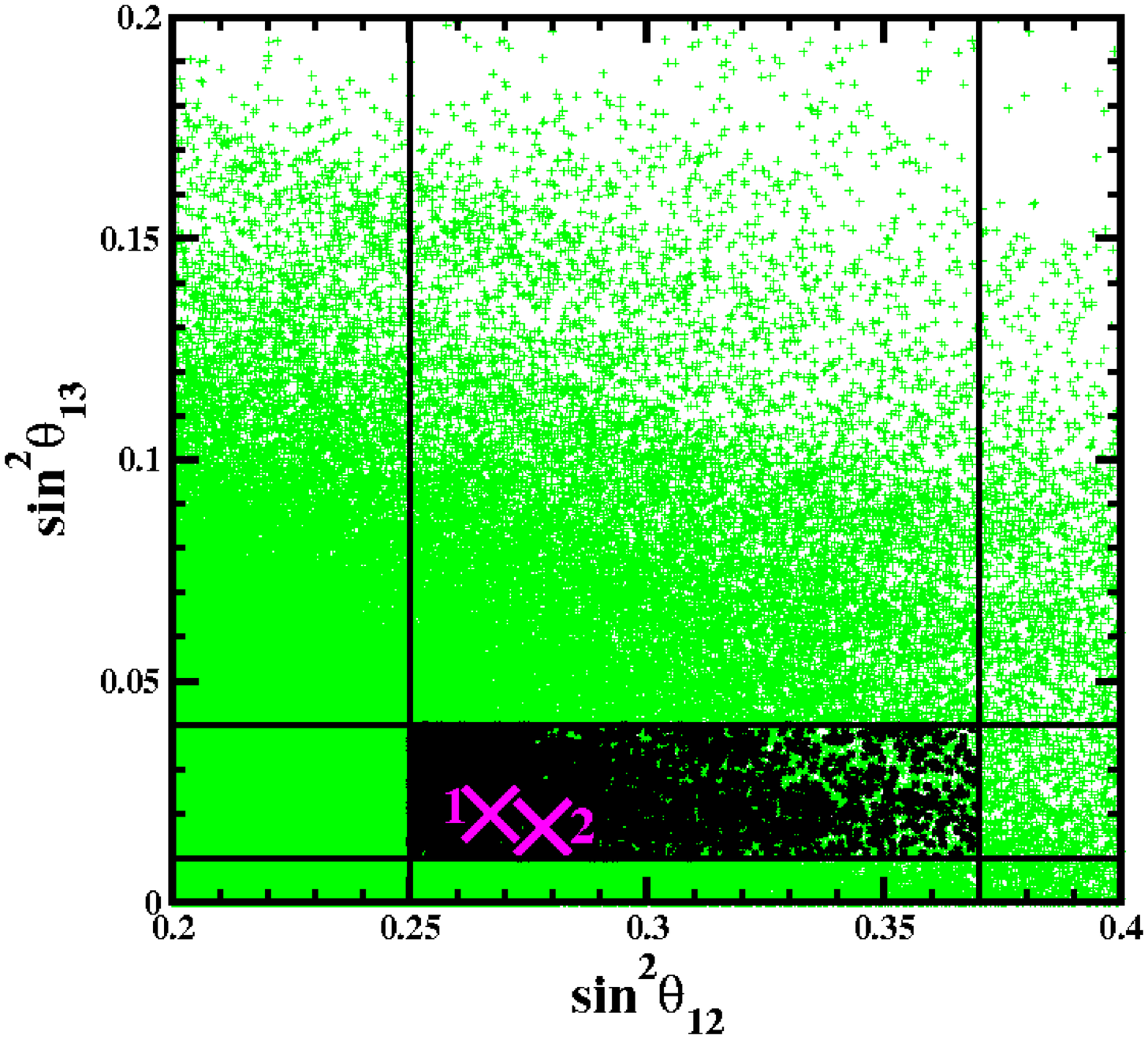}
\includegraphics*[scale=0.43]{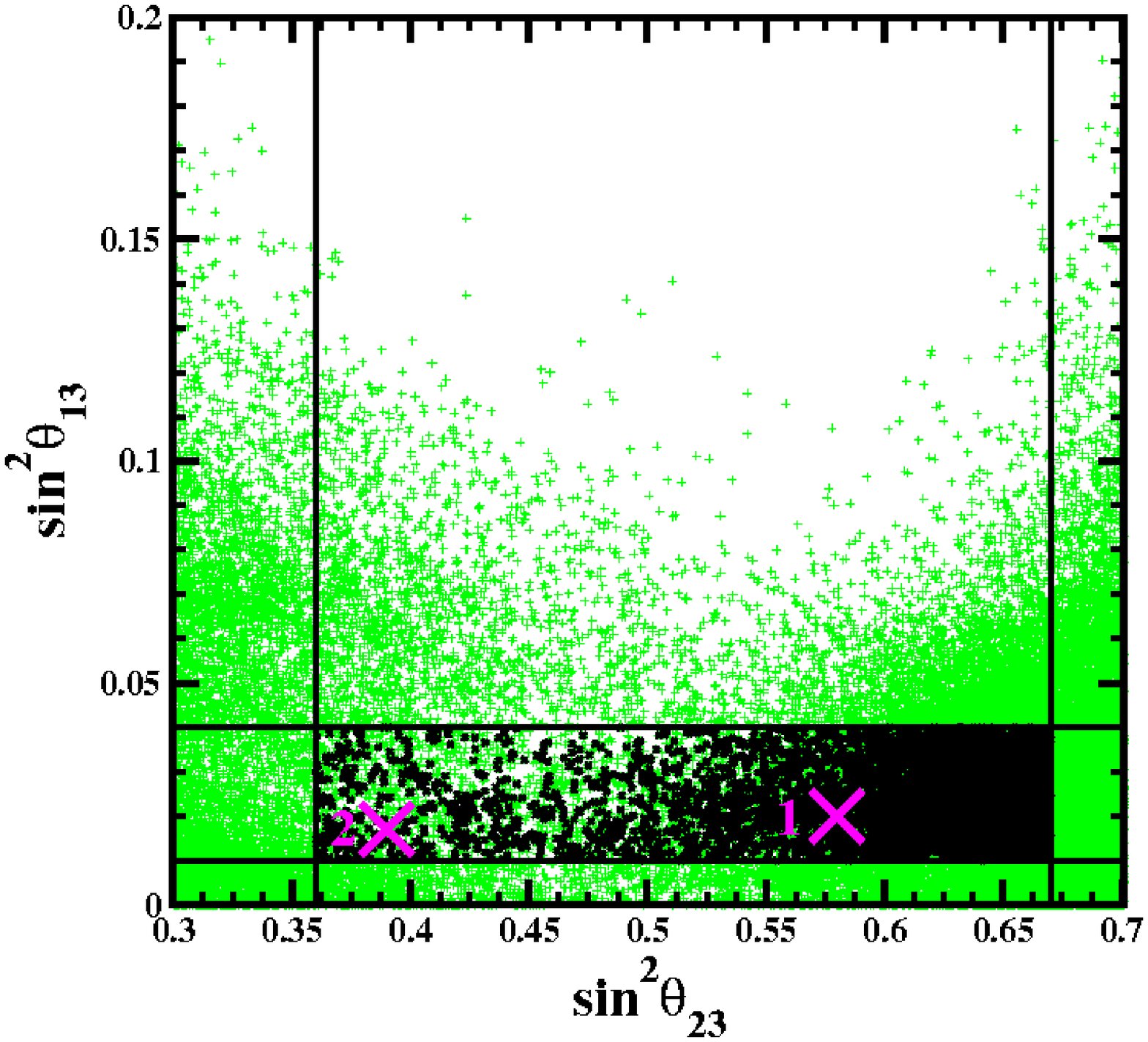}
\end{center}
\caption{\it We summarise the correlations between the neutrino mixing 
angles before and after constraining the model coefficients (as
discussed in the text). The solid lines 
indicate the experimental bounds, and the small black crosses represent models satisfying all 
constraints. The two large magenta crosses 
correspond to the benchmark models that are discussed in the text and 
in Table~\ref{tab:table1} and
are used for numerical
calculations.}
\label{fig:t13t23_nat}
\end{figure*}

In Fig.~\ref{fig:t13t23_nat} we present the predictions for the neutrino mixing 
angles corresponding to the above criteria. 
Here we do not take into account the RGE run of the mass matrices,
since these effects will not affect the global picture of the  solutions displayed in Fig.\ref{fig:t13t23_nat}.
Within this class of models, most of the solutions that
reproduce the correct range of  $\theta_{12}$ and
$\theta_{23}$, also predict a
neutrino mixing angle $\theta_{13}$ that is compatible with the
data from~\cite{daya-bay,reno}. 
Within the range of $\theta_{13}$ the model predictions are mostly in the
mid-lower range of $\theta_{12}$. On the other hand, in the case of $\theta_{23}$,
a higher density of solutions is  found in the mid-higher range of $\theta_{23}$. 
We note that the number of fits predicting the maximal value  $\theta_{23}=\pi/4$  
is smaller when we impose the hierarchy condition $(iii)$, as compared
to the case 
where only the experimental bounds on $\Delta^2 m_{\text{sol}}$ and  
$\Delta^2 m_{\text{atm}}$ are considered. This is consistent with the  observed deviation of 
$\theta_{23}$ from its maximal value~\cite{neu-fits}. 

As benchmarks for studying LFV in subsequent sections,
we have identified the two sets of textures of Table~\ref{tab:table1}, and
indicated with crosses in Fig.~\ref{fig:t13t23_nat}.
These benchmarks are chosen as representatives of the two different regions identified in the 
global statistical data analysis performed in~\cite{neu-fits} and also
shown in  Fig~\ref{fig:t13t23_nat}.
Fit 1 lies in the right region of the right panel of Fig~\ref{fig:t13t23_nat}, with larger  $\theta_{23}$,
and Fit 2 lies in the left region of the same panel, with smaller $\theta_{23}$.  

We would like to point out that the fits in
Fig.~\ref{fig:t13t23_nat} are independent of the charges $n_i$, which  
affect $Y_\nu$ and $M_N$ but not their combination in $m_{eff}$. On the other hand, 
the choices of $n_i$ do affect the rates for charged LFV processes, as we will show in the next Section.
Charged LFV processes are therefore powerful probes of parameters that cannot be constrained by lepton 
hierarchies and better measurements of the neutrino parameters.

\section{Charged-Lepton-Flavour Violation}

\subsection{Slepton masses in see-saw models.}
\label{sec:3.1}

The neutrino mass textures discussed above arise naturally from $SU(5)$ enhanced by a U(1) flavour symmetry. 
In order to study LFV processes, however, we are guided to a large extent by phenomenological considerations and thus our results are more generic. The benchmark solutions previously identified, which naturally reproduce the
correct neutrino phenomenology, are used to obtain the matrices that determine the LFV vertices 
in the context of  the CMSSM  (extended with right-handed heavy Majorana neutrinos arising from a see-saw mechanism).

Since the Dirac neutrino and charged-lepton Yukawa
couplings cannot, in general, be diagonalized simultaneously (and
since both types of lepton Yukawa
couplings appear in the RGEs) the lepton Yukawa
matrices and the slepton mass matrices at low energies cannot be
diagonalized simultaneously, either.
In the basis where the charged-lepton masses $m_{\ell}$
are diagonal, the soft slepton-mass matrix acquires 
corrections that contain off-diagonal contributions from renormalization at scales below
$M_{GUT}$, of the following form in the leading-log approximation~\cite{LFVhisano}:
\begin{equation}
\delta{m}_{\tilde{\ell}}^2\propto \frac 1{16\pi^2} (6m^2_0 + 2A^2_0)
{Y_{\nu}}^{\dagger} Y_{\nu}  \ln(\frac{M_{GUT}}{M_N}) \, ,
\label{offdiagonal}
\end{equation}
Here $M_N$ is the intermediate scale where the effective neutrino-mass operator is formed. 
The physical charged slepton masses are obtained by numerical
diagonalization of the following $6 \times 6 $ matrix:
\begin{equation}
\label{eq:66}
{m}_{\tilde{\ell}}^2=\left(
\begin{array}{cc} 
m_{LL}^2&m_{LR}^2\\
m_{RL}^2&m_{RR}^2 
\end{array}\right) \, ,
\end{equation}
where all the entries are $3 \times 3$ matrices in flavour space.
Using the basis where $Y_{\ell}$ is diagonal, it is convenient to 
write the $3\times 3$ entries of (\ref{eq:66}) in the form:
\begin{eqnarray}
m_{LL}^2&=& (m_{\tilde{\ell}}^\text{diag})^2+\delta{m}_{\tilde{\ell}}^2 
+m_{\ell}^2 - \frac{1}{2}(2 M_W^2  -M_Z^2) \cos 2 \beta \, , \\
m_{RR}^2&=& (m_{\tilde{\ell}_R}^\text{diag})^2+m_{\ell}^2-
(M_Z^2-M_W^2) \cos 2\beta \, ,\\
m_{RL}^2&=& (A_\ell^\text{diag} +\delta A_\ell - \mu \tan\beta) m_{\ell} \, ,\\
m_{LR}^2&=& m_{RL}^{2\dagger} \, .
\label{bits}
\end{eqnarray}
Here  $\tan\beta$ is the ratio of the two MSSM Higgs vevs,
$(m_{\tilde{\ell}}^\text{diag})^2, (m_{\tilde{\ell}_R}^\text{diag})^2$ and $A_\ell^\text{diag}$ 
denote the diagonal contributions to the corresponding matrices,
obtained by numerical integration of the RGEs, and 
$\delta{m}_{\tilde{\ell}}^2$ and $\delta A_{\ell}$
denote the corrections to off-diagonal terms that appear because
$Y_{\nu}$ and $Y_{\ell}$ cannot be diagonalized simultaneously.

The  full mass matrix for left- and right-handed 
sneutrinos has a $12\times 12$ structure,
given in terms of $3\times 3$
Dirac,  Majorana  and sneutrino mass matrices.
The effective $3  \times 3$
mass-squared matrix for the left-handed sneutrinos has the same
form as the $m_{LL}^2$
part (\ref{bits}) of the $6 \times 6$ charged-slepton matrix
(\ref{eq:66}), with the difference that now the Dirac masses
are absent. 
In Ref.~\cite{gllv} it was shown that is sufficient to use
\begin{equation}
{m}_{\tilde{\nu}}^2= (m_{\tilde{\ell}}^\text{diag})^2+ \delta m_{\tilde{\ell}}^2 + \frac{1}{2}
M_Z^2 \cos 2\beta  \, .
\end{equation}

The matrix responsible for LFV in the lepton-slepton-gaugino vertices is
\begin{equation}
V_{LFV}=V_D^\dagger V_\ell
\label{VLFV}
\end{equation}
and the slepton mass matrices contain
off-diagonal terms generated by: 
\begin{equation}
m_{LL}^2=V_{LFV}^\dagger (m_{LL}^2)_{\text{d}} V_{LFV} \, ,
\label{eq:mLL}
\end{equation}  
while the $A$-terms become:
\begin{equation}
A_\ell=V_{LFV}^T (A_\ell)_\text{d} \, .
\label{eq:A}
\end{equation}  
Here $(m_{LL}^2)_{\text{d}}$ and  $(A_\ell)_{\text{d}}$ are the terms resulting from 
the RGE running of the universal soft terms at the GUT scale in a basis where $Y_\nu$ is diagonal. 
The corresponding effects in $m_{RR}^2$ are negligible 
and are not considered in the numerical calculations.

We remark that, in general, in the framework of susy SU(5) GUT with U(1) family symmetries, flavour 
dynamics are linked to scalar singlet fields, flavons, whose non-zero vacuum expectation 
value breaks the U(1) symmetry. The RGE running of the parameters above the GUT scale
due to flavons dynamic induces flavour dependent corrections to sfermion soft mass matrices and A-terms 
and thus potentially large LFV
effects~\cite{gllv,Babu,Olive,Kane,Altarelli}. However, while flavon
effects can be potentially very large, we know from flavour
phenomenology that this is not the case and that they have to be
suppressed to the point that they are comparable to the effects we
consider here. Such a suppression 
can be achieved, among others, in an scenario where
the effect of non-universal soft terms is diminished by RGE effects
beyond the GUT scale, as indicated in \cite{pedro}.
Given that the exact knowledge of flavon effects depends on model building and physics of unknown scales (supersymmetry breaking scale, flavon dynamics scale
or string scale), a complete mechanism cancelling the undesired
flavour-violating soft terms at $M_{GUT}$ goes beyond the scope of
this paper.

\subsection{LFV and neutrinos }
\label{sec:3.2}

We now study the conditions under which the favoured range of 
neutrino masses and mixing, can lead in a natural way
to observable signatures for charged LFV. 
As already discussed, while the neutrino parameters are independent of
the charges $n_i$, this is not the case for LFV. As a result, 
LFV can provide a way to probe the right-handed neutrino sector, for
which only limited information is available. 

In section~\ref{sec:2.1}, we identified two representative benchmark fits 
suitable for studying charged LFV.
The level of charged LFV is determined by  the product
$V_{LFV}=V^\dagger_D V_{\ell}$, and thus by the
charges $n_i$ which enter in the Dirac neutrino mixing 
matrix $V^\dagger_D$. 
Having a $V_{\ell}$ with large off-diagonal 1-2 and 2-3 entries is a
natural choice to match the lepton data. Then, different choices of
$n_i$ lead to different possibilities for $V^\dagger_D$; in fact,
there are two possibilities associated with 
a $V_{\ell}$ with large off-diagonal elements: (1)
the charge combinations generate a $V^\dagger_D$ with small
off-diagonal elements;
in this case, $V_{LFV} \sim V_{\ell}$.
(2) the off-diagonal elements of $V^\dagger_D$ are large, but
multiplied 
with $V_{\ell}$, they can give either large or small elements in $V_{LFV}$ depending on 
coefficients and phases. 

An illustration of the dependences of the entries in $V_{LFV}$ on the different
right-handed neutrino charges is given in Table~\ref{tab:table2}. 
For simplicity, we focus on Fit 2, noting that similar results hold for Fit 1.
The matrix $(i)$ is an example of  case (1), with  small off-diagonal elements in $V_D$.
In $(ii)$ and $(iii)$, $V_D$ has large off-diagonal elements which enhance $V_{LFV}$.
Finally in $(iv)$,  $V_D$ has large off-diagonal elements but cancellations with $V_{\ell}$ occur
in the 2-3 sector, suppressing LFV.
\begin{table}[t!]
\begin{center}
\resizebox{\textwidth}{!}{
\begin{tabular}{ccccc}
\hline\hline
\\
      &  (i)        & (ii)    &(iii)   &    (iv)   \\
$n_i$ &  $\{n_1=1,n_2=0,n_3=0\}$ &  $\{n_1=2,n_2=1,n_3=0\}$ &
  $\{n_1=2,n_2=0,n_3=1\}$ &  $\{n_1=0,n_2=1,n_3=0\}$ \\
\\
$V_{LFV}$ &
$\left(\begin{array}{rrr}
0.805 & -0.385 & -0.451 \\
0.182 & 0.885 & -0.429 \\
0.565 & 0.263 & 0.782
\end{array}\right)$ &
$\left(\begin{array}{rrr}
0.805 & -0.385 & -0.452\\
-0.064 & 0.700& -0.711 \\
0.590 & 0.601 & 0.539
\end{array}\right)$ & 
$\left(\begin{array}{rrr}
-0.805 & 0.384 &0.453\\
0.544& 0.782 & 0.305 \\
-0.237 & 0.492 & -0.838
\end{array}\right)$ &
$\left(\begin{array}{rrr}
0.806& -0.401 & -0.436 \\
-0.437 & -0.899 & 0.016 \\
-0.399 & 0.178 & -0.901
\end{array}\right)$ \\
\\
\hline\hline
\end{tabular} }
\end{center}
\caption{\it Values for the matrix $V_{LFV}$ of  Eq.~(\ref{VLFV})
corresponding to Fit 2 of Table~\ref{tab:table1} with $\varepsilon=0.2$ and different choices of the 
U(1) charges, Eqs.~(\ref{Yukawas},~\ref{mN}). }
\label{tab:table2}
\end{table}

\subsection{Numerical procedure and RGEs}
\label{sec:3.3}

The recent LHC measurement of the Higgs mass~\cite{higgs1,higgs2} 
imposes severe constraints in the CMSSM parameter space.  
More specifically, a Higgs masses of $m_h \sim125$~GeV implies, in general, 
a relatively heavy sparticle spectrum, which is consistent with the cosmological constraint
on the neutralino relic density only in limited regions.
A global analysis of the CMSSM parameter space was performed in~\cite{Buchmueller:2012hv},
yielding two almost equally good fits to the available data, one with relatively low
sparticle masses and $\tan \beta \sim 16$, and the other with larger sparticle masses
and $\tan \beta \sim 45$~\footnote{Note that our $A_0$ values have
  opposite sign with respect to
those of Ref.~\cite{Buchmueller:2012hv} where the authors use a definition
for the trilinear scalar coupling that differs from the one in standard codes like Suspect
and SoftSusy.}:
\begin{eqnarray}
&(a)& \tan\beta=16,\;\;\; m_0=300 \rm{~GeV},\;\;\; M_{1/2}=910 {\rm ~GeV}, \;\;\; 
A_{0}=1320 {\rm ~GeV} \, , \nonumber  \\
&(b)& \tan\beta=45,\;\;\; m_0=1070 \rm{~GeV},\;\;\; M_{1/2}=1890 {\rm ~GeV}, \;\;\; 
A_{0}=1020 {\rm ~GeV} \, . 
\label{benchmarks}
\end{eqnarray}
The  sign of $\mu$ is positive, as favored by $g_\mu-2$ measurements.
Regarding cosmological considerations, point 
$(a)$  belongs to  the area where the WMAP-favoured range of $\Omega_{\chi} h^2$ is obtained via 
$\chi-\tilde{\tau}$ coannihilation~\footnote{We
note that in this region the $\chi-\tilde{\tau}$ mass difference is very small, offering other experimental 
challenges and opportunities~\cite{CELMOV,CannoniLHC}.}, whereas point $(b)$
lies in the funnel region where the neutralino LSP annihilates rapidly via direct-channel $H/A$ poles.

We evaluate the RGEs using universal soft terms at the  GUT scale, $M_{GUT}$.
The standard model parameters are eveluated at $M_Z$ and $m_t(m_t)$. 
At the GUT scale, defined as the meeting point of the gauge couplings 
$g_1$ and $g_2$ ($g_3$ is set 
so that $\alpha_s(m_Z)=0.1172$),
we work in a basis where $Y_\nu$ is diagonal. Non-diagonal elements of the soft mass matrices  
are induced from the fact that $Y_{\ell}$ cannot be diagonalized simultaneously with $Y_\nu$.  
The right handed neutrino scale is identified with the mass of the largest eigenvalue of $M_N$, $M_3$.  
The coupling $Y_{\nu_3}$ is calculated by requiring that $m_{\nu_3}=0.05$~eV at low energy, 
using the respective RGE~\cite{Ellis:1999my,Antusch,Casas}.
At $M_3$ we decouple  the see-saw parameters from the RGE; in doing so, we neglect the effect of the 
lighter neutrinos, which in the case of hierarchical neutrinos is  not
large
(even if $M_2$ and $M_3$ are much 
lighter than $M_3$, the corresponding $Y_\nu$ must decrease according to the see-saw relation, resulting to an 
insignificant impact on the slepton mass running).
At $M_3$ we rotate all the fields in the basis where $Y_{\ell}$
becomes diagonal; in this basis, $m_{LL}^2$ and $A_{\ell}$ take 
the form of Eqs.~(\ref{eq:mLL}) and ~(\ref{eq:A})  while $m_{RR}^2$ remains essentially  diagonal 
since its RGE is not affected by $Y_\nu$. Moreover, in this basis, only the diagonal terms evolve from 
$M_3$ down to low energies. 
The matrix $V_{LFV}$ is computed using Yukawa textures that match the
neutrino data. In our RGE analysis we do 
take into account the change of the overall 
scale of $m_{eff}$ and $Y_{\ell}$ but not  the  RGE dependence of each matrix 
element (for hierarchical neutrinos this dependence is 
small and can be absorbed in the uncertainty of the coefficients used to fit the texture 
without a significant effect in the slepton mass matrices).
At low energies, we decouple the SUSY particles at $M_{SUSY}=\sqrt{m_{\tilde{t}_1}\cdot m_{\tilde{t}_2} 
}$ and continue with the SM RGEs to $m_t$ and $M_Z$, with the initial conditions of the RGEs being iteratively 
adjusted to the experimental data.

\subsection{Predictions for Radiative Decays}
\label{sec:3.4}

\begin{figure*}[tp!]
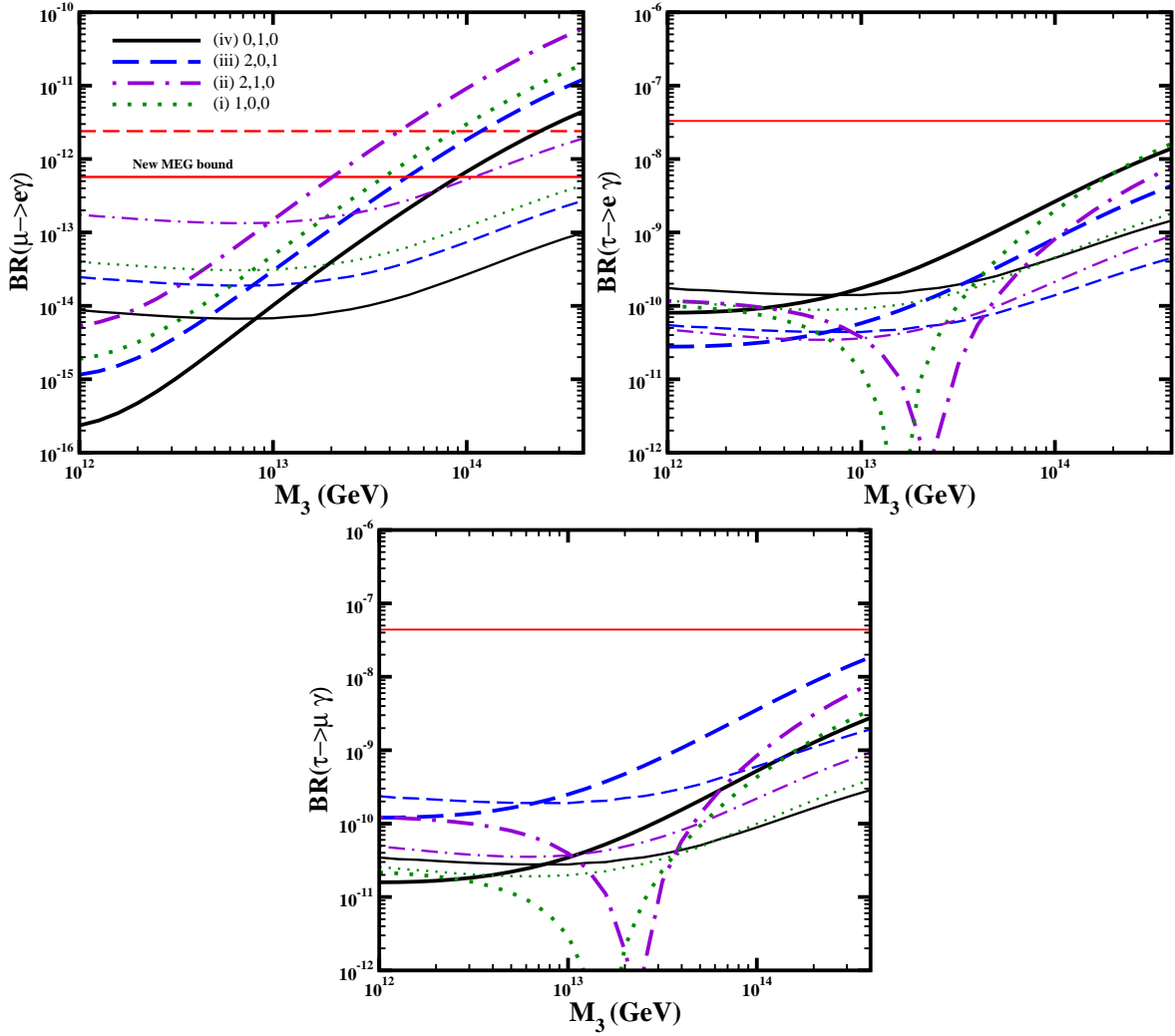

\begin{center}
\includegraphics*[scale=0.4]{bmegrh.eps}
\includegraphics*[scale=0.4]{btegrh.eps} 
\includegraphics*[scale=0.4]{btmgrh.eps}
\end{center}
\caption{\it Predictions for the rare LFV decays 
$\ell_i\rightarrow \ell_j \gamma$ as a function of  the right-handed neutrino mass
$M_N$, for the benchmark points displayed in (\ref{benchmarks}) $(a)$
(thick line), $(b)$ (thin line),
using the neutrino mixing fits shown in Table \ref{tab:table2}.  
The solid lines correspond to case $(iv)$, dashed ones to $(iii)$, 
dot-dash to $(ii)$ and dot to $(i)$ . The horizontal solid 
lines indicate the current experimental upper bounds,
 while the dashed line correspond to the previous MEG limit on $BR(\mu\rightarrow e \gamma)$.}
\label{fig:brrh1}
\end{figure*}

The matrix element of the electromagnetic-current operator between two distinct lepton mass eigenstates 
$\ell_i$ and $\ell_j$ is given in general by
\begin{eqnarray}
{\cal T}_\la &=& \langle \ell_i|(p-q)|{\cal J}_\la|\ell_j(p)\rangle\nonumber\\
{  }&=&{\bar u_i}(p-q)
      [ m_j i\sigma_{\la\beta}q^\beta
               \left(A^L_MP_L+A^R_MP_R\right)+\nonumber\\
{  }&&\quad\quad    (q^2\gamma_\la-q_\la\gamma\cdot q)
               \left(A^L_EP_L+A^R_EP_R\right)] u_j(p) \, ,
\label{general}
\end{eqnarray}
where $q$ is the photon momentum. The 
coefficients $A_M$ and $A_E$ denote 
contributions from neutralino/charged-slepton 
and chargino/sneutrino exchanges.
The amplitude of the LFV process is then proportional to ${\cal T}_\la
\epsilon^\la$, where $\epsilon^\la$ is the photon-polarization vector.
The branching ratios (BR) of the decays $\ell_j \ra \ell_i+\gamma$ are
calculated using the exact formulas of Ref.~\cite{LFVhisano}. 

In Fig.~\ref{fig:brrh1} we show numerical predictions 
for the LFV branching ratios arising from the textures introduced above 
and for the CMSSM parameters specified in (\ref{benchmarks}).
We show the effect of varying $M_3$ from $6 \times
10^{14}$~GeV down to $10^{12}$ GeV, for Fit 2 of Table~\ref{tab:table1}
and for the choices of right-handed neutrino charges of Table~\ref{tab:table2}.
We can see that the experimental upper bounds on
BR($\mu\rightarrow e \gamma$) can be reached  
with some of the  Yukawa textures we studied, 
even with the heavy sparticle spectrum implied by the benchmark point $(a)$.
The new MEG bound on BR($\mu\ra e \gamma$)  imposes constraints on the see-saw scale 
for all  charge choices of Table~\ref{tab:table2} for 
point $(a)$, and for fit $(ii)$ at point $(b)$. 

In the case of point $(a)$, we find cancellations that reduce the branching ratios for some neutrino mixing 
fits. This happens because of the large value of $A_0$, which leads to significant cancellations among the 
different LFV decay amplitudes~\cite{LFVhisano, gllv,Mismatch}.  For the fits $(i)$ and $(ii)$ we find this 
type of cancellation in BR($\tau\ra \mu \gamma$) and  BR($\tau\ra e \gamma$), for a 
range of values of the right-handed neutrino scale $M_3$. This is due to the fact that, at large values of 
$A_0$, the contribution $A^R_M$ from the neutralino/charged slepton loops in eq.~(\ref{general}) 
cancels with the one arising from chargino/sneutrino loops, $A^R_E$, which is the dominant 
contribution for small $A_0$. 
The ratio of these two contributions can be modulated by the parameters 
that determine the size of the flavor mixing elements $\delta{m}_{\tilde{\ell}}^2$ in 
eq.~(\ref{bits}). In our case the scale of $M_3$  also determines 
the strength of $Y_\nu$ and thus the size of the LFV terms.  
We use  Fig.~\ref{fig:brrh1} and the current MEG bound on BR($\mu \ra e \gamma$) to fix 
the $M_3$ scale for further studies: $M_3=2\times 10^{13}$~GeV for the
benchmark point $(a)$ 
and $M_3=10^{14}$~GeV for the benchmark point $(b)$.

\subsection{LFV in $\chi_2$ decays at the LHC}
\label{sec:3.5}
 
A promising channel to search for LFV at the LHC
is the production and decay of the second lightest neutralino,
$\chi_2 \to \chi + \tau^\pm \mu^\mp$.
In~\cite{LHCh,CEGLR2} it was shown that in order to have a signal that 
could be distinguished from the background, the ratio
\begin{equation}
R_{\tau \mu}=\Gamma(\chi_2\rightarrow\chi+ \tau^\pm+ \mu^\mp)/
\Gamma(\chi_2\rightarrow\chi+ \tau^\pm+ \tau^\mp)
\label{eq:ratio}
\end{equation}
should be of the order of 10\%. For $A_0 =0$, due to the absence of
cancellations suppressing rare charged lepton decays, one had to go beyond the
CMSSM to find solutions compatible with all experimental and
cosmological data \cite{CEGLR2}.
Here, we extend this study to large values of $A_0$,
noting that the cancellations that can arise in the branching ratios 
of radiative decays do not 
occur in $R_{\tau\mu}$. This opens the possibility to observe  
LFV  in neutralino decays at the LHC, in cases where 
LFV would be undetectable in rare charged lepton radiative decays.

To see whether this is indeed the case, 
we proceed with the computation 
including all contributing on-shell sfermion exchange diagrams, 
as given in~\cite{Bartl}:
\begin{equation}
\text{BR}(\chi_2\rightarrow\chi \tau^\pm \mu^\mp)=\sum_{i=1}^3 \left[
\text{BR}(\chi_2\rightarrow\tilde{\ell}_i \mu)\text{BR}(\tilde{\ell}_i\rightarrow\tau \chi) + 
\text{BR}(\chi_2\rightarrow\tilde{\ell}_i \tau)\text{BR}(\tilde{\ell}_i\rightarrow\mu \chi)
\right] 
\label{eq:R}
\end{equation}
These are evaluated in the benchmark points 
$(a)$ and $(b)$, to see whether
the branching ratio  can be of  the order of the required reference value.
 
\begin{figure}[t!]
\begin{center}
\includegraphics*[scale=0.40]{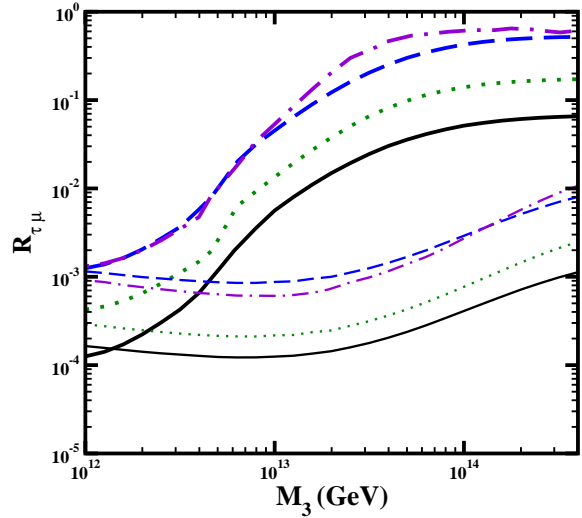}
\end{center}
\caption{\it   
The ratio defined in Eq.~(\ref{eq:ratio}) is presented for the CMSSM points $(a)$
(thick line) and $(b)$ (thin line)
(in Eq.~\ref{benchmarks}), with the
same notation as in Fig.~\ref{fig:brrh1}.
}
\label{fig:HP}
\end{figure}

In Fig.~\ref{fig:HP} we present the predictions for the
branching ratio~(\ref{eq:ratio}) as a function of $M_3$. For point
$(a)$, our predictions are within the reach of the LHC 
for values of $M_3$ that are compatible with the MEG limit.
For point $(b)$, the predictions are below the expected experimental sensitivity.

\subsection{LFV  at a Linear Collider}
\label{sec:3.6}

In supersymmetric models where LFV is produced by lepton-slepton
vertices, observable signatures may occur either directly, in slepton-pair production,
or  indirectly, via slepton production in cascade decays~\cite{sleptonscemu}. 
If the flavour mixing is introduced in the left-left slepton
sector, as is the case for the models under consideration here, the
dominant
channels are slepton-pair production and LFV decays, such as: 
\begin{eqnarray}
e^+ e^- & \rightarrow & \tilde{\ell}_i^{-} \tilde{\ell}_j^{+} \rightarrow 
\tau^{\pm} \mu^{\mp} \tilde{\chi}^{0}_1 \tilde{\chi}^{0}_1 , \nonumber \\
e^+ e^- & \rightarrow & \tilde{\nu}_i \tilde{\nu}_j^c  \rightarrow 
\tau^{\pm} \mu^{\mp} \tilde{\chi}^{+}_1 \tilde{\chi}^{-}_1 \, .
\label{eq:pair}
\end{eqnarray}
In the CMSSM benchmark points introduced above, 
the channel mediated by charged sleptons
clearly dominates over the sneutrino-pair production process, and may  lead to a cross section of the 
order of 1 fb; this is the reference value of \cite{CEGL-LC}, for a 
LFV signal of $\mu^\pm\tau^\pm$ pairs that can be distinguished from the 
background, according to the study made in ~\cite{LFV-LC2}. 
Here, we extend our previous results~\cite{CEGL-LC} which were focused on the production of $\mu^\pm\tau^\pm$ pairs  by considering the full structure of the 
Yukawa matrices, thus comparing the 
LFV production of charged leptons of all generations. 
Complete expressions for the LFV cross sections are
given in Ref.~\cite{LFV-LCa} and used in our work. 
\begin{figure*}[t!]
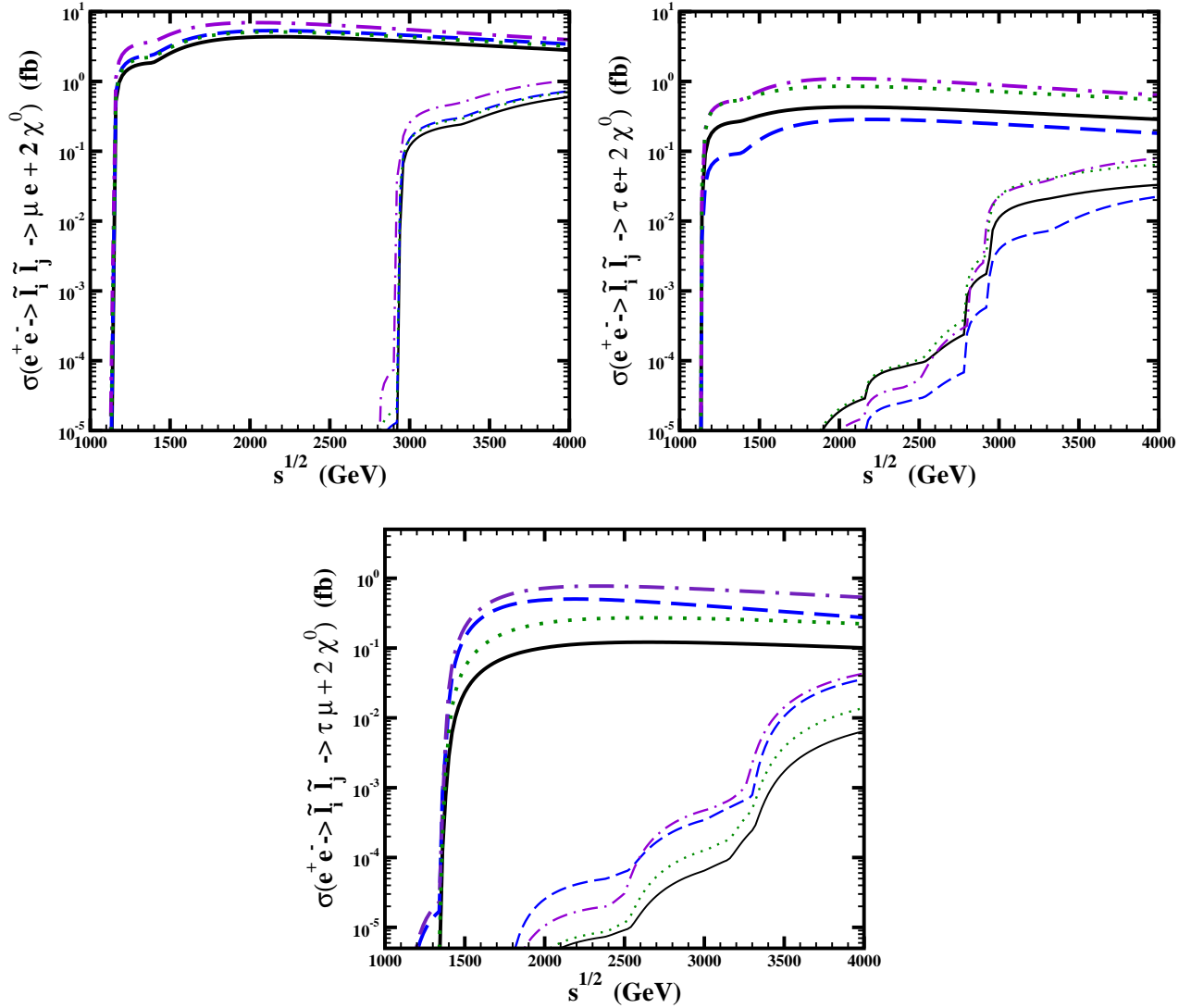

\begin{center}
\includegraphics*[scale=0.41]{cme_s.eps}
\includegraphics*[scale=0.41]{cte_s.eps}
\includegraphics*[scale=0.41]{ctm_s.eps}
\end{center}
\caption{
\it Values of the cross sections $\sigma(e^+e^-\rightarrow 
\tilde{\ell}_i^{-} \tilde{\ell}_j^{+} \rightarrow \ell_a ^\pm \ell_b^\mp +2 \chi^0)$ 
($\ell_a\neq \ell_b$ as indicated in each panel) as functions of $\sqrt{s}$. 
The line styles are the same as those in
Fig.~\ref{fig:brrh1}. For point $(a)$  we use $M_3=2 \times 10^{13}$~GeV, while for the point $(b)$  
we work with $M_3= 10^{14}$~GeV.}
\label{fig:css}
\end{figure*}

In Fig.~\ref{fig:css} we present the expected cross sections 
$\sigma(e^+e^-\rightarrow 
\tilde{\ell}_i^{-} \tilde{\ell}_j^{+} \rightarrow \ell_a ^\pm \ell_b^\mp +2 \chi^0)$
as a function of $\sqrt{s}$ for the same choice of parameters as in Figures~\ref{fig:brrh1} 
and~\ref{fig:HP}.
Naturally, the cross sections in the case of point $(a)$ are larger because sleptons and
gauginos are much lighter than in the spectrum of point $(b)$. 
In $(a)$ sleptons 
and sneutrinos are nearly degenerate and the cross sections, at energies above the threshold for 
pair production that is around $1.2$ TeV, show a feeble decrease with $\sqrt{s}$. Final states
with $e\mu$ pairs have the largest cross section, with value between 1
fb and 10 fb, with a small
dependence on the choice of charges (i)-(iv). On the other hand, the cross section for the 
processes with $\tau e$ and $\tau\mu$ final states show a stronger
dependence on the choice of 
 charges, 
varying between $10^{-1}$ fb and 1 fb in the first
case, and between $10^{-2}$ 
and 1 fb in the second case.
Similar behaviour is observed in the case of point $(b)$ where the heavy spectrum implies 
a threshold around 3 TeV and cross sections below $10^{-1}$ fb.

According to Fig.~\ref{fig:brrh1}, at the selected value of  $M_3=2 \times 10^{13}$~GeV, 
BR($\tau\ra\mu \gamma $) and  BR($\tau\ra e \gamma $) are 
suppressed.  Since these cancellations do not occur for the LFV LC signals, it is possible to observe slepton flavor oscillations at the LC, in cases where 
LFV would be undetectable in rare charged lepton decays (as it could also happen at the LHC).
It is worth to remark that the CLIC project for a linear collider 
has as nominal centre of mass energies the values 1.4 TeV and 3 TeV~\cite{CLIC,Battaglia},
with the option of reaching 5 TeV. The value $\sqrt{s}=1.4$ TeV is optimal
for point $(a)$ where the LFV cross sections are nearly maximal. 

\section{LFV and Leptogenesis}
\label{sec:4}

We comment now on possible
links between our LFV predictions and leptogenesis~\cite{leptogenesis} through the
decays of heavy, right-handed Majorana neutrinos into leptons and
antileptons. Since LFV is related to the see-saw parameters  in our framework,  there can be 
interesting consequences for LFV in charged lepton
decays and elsewhere~\cite{leptogenesis-LFV}.

In previous Sections, we have used real parameters to fit the Yukawa couplings, 
but small phases that would not alter our LFV
considerations could induce significant contributions to the lepton
and baryon asymmetries of the universe. In what follows, we
explore what sizes of the phases in $Y_\nu$ 
can predict a value for the baryon asymmetry $Y_B$ 
compatible with the observation \cite{WMAP-bar}
\bea
Y_B = (6.16 \pm 0.16) \times 10^{-10} \, .
\label{eq:ybexp}
\eea
For hierarchical heavy neutrinos in a supersymmetric see-saw model, 
one has~\cite{Giudice},
\bea
Y_B\simeq- 10^{-2} \kappa \epsilon_1 \, ,
\eea
where $\epsilon_1$ is the CP-violating asymmetry in the decay of the
lightest Majorana neutrino and $\kappa$ an efficiency factor parametrizing the
level of washout of the generated asymmetry by
inverse decay and scattering interactions. The latter
depends on
the mass of the decaying neutrino
$M_1$ 
and the effective mass parameter
\bea
\tilde{m}_1={v_u^2\over M_1}(\lambda_\nu^\dagger \lambda_\nu)_{11} \, ,
\eea
where  $\lambda_\nu$ is the Dirac neutrino Yukawa  matrix in the basis where
the Majorana masses are diagonal, and $v_u$ is the vev of the Higgs field
that couples to up-quarks and neutrinos.

The CP-violating decay asymmetry $\epsilon_1$ arises from the
interference between tree-level and one-loop amplitudes:
\bea
     \epsilon_1 & =  & 
{1\over( 8 \pi \lambda_{\nu}^{\dagger}\lambda_{\nu})_{11}}
     \sum_{i\neq 1} {\rm Im} 
\left [
((\lambda_{\nu}^{\dagger} \lambda_{\nu})_{1i})^2
\right ] f \left 
(\frac{M^2_{1}}{M^2_{i}} \right ) \, , 
\label{eq:eps}
\eea
with $f(y)  =  \sqrt{y}\left[{1\over 1-y} + 1-(1+y)\ln\left({1+y\over y}
     \right)\right]$.
The value of the 
CP asymmetry depends
on the details of the model, but
a model-independent upper bound exists, given
by~\cite{DI}
\begin{equation}
|\epsilon_1| \leq \frac{3}{8 \pi} \frac{M_1}{v_u^2} (m_3 - m_1) \, ,
\label{eq:ebound}
\end{equation}
where the $m_i$ are the masses of the light neutrinos.

In our work, the flavour and GUT symmetries enable us
to correlate $M_1$ with the mass of the heaviest
right-handed neutrino, $M_3$, which  
controls the LFV effects. The neutrino Yukawa couplings
of all generations are also related.  Then, using as a first approximation 
eq. (\ref{eq:ebound}) and  $\kappa$ as derived in~\cite{Giudice}, we can infer
how leptogenesis 
may be accommodated in our study. Our fits predict large $\tilde{m}_1$
that, according to \cite{Giudice}, implies a strong
 wash-out  regime in which $\kappa$ ranges 
between $\sim 10^{-3}$ and $\sim 10^{-4}$. 

Some typical results are presented in Table~\ref{tab:table3}, where we see that 
at a reference value of $M_3=5\cdot 10^{13}$~GeV, 
$Y_B^{max}$ (calculated using $\epsilon_{1}$ from eq. (\ref{eq:ebound})) 
is considerably larger than the experimental value of $Y_B$ for fit $(i)$,
implying that in this case the CP-violating phases have to be
small enough for $\epsilon_1$ to be
well below its maximal value. On the other hand, $Y_B^{max}$ is below $Y_B$ for 
fit $(iii)$, and of the same order of magnitude for fits $(ii)$ and
$(iv)$. These differences between the fits are
due to the different hierarchies between the heavy Majorana masses
and the neutrino Yukawa couplings in $\tilde{m}_1$ 
(which are determined by the right-handed
neutrino charges). Consequently, they
indicate how leptogenesis can be used as an additional probe of
the right-handed neutrino sector, for which very limited information is
provided by the neutrino data alone. 

\begin{table}[t!]
\begin{center}
\begin{tabular}{|c|c|c|c|c|}
\hline\hline
      &  (i)        & (ii)    &(iii)   &    (iv)   \\
\hline
$M_1 (GeV)$ &
$\begin{array}{r}
4.3 \cdot 10^{12}\\
8.6 \cdot 10^{10}
\end{array}$ &
$\begin{array}{r}
2.6 \cdot 10^{11}\\
5.3\cdot 10^{9}
\end{array}$ &
$\begin{array}{r}
5.4 \cdot 10^{11}\\
1.1 \cdot 10^{10}
\end{array}$ &
$\begin{array}{r}
2.3 \cdot 10^{12}\\
4.7 \cdot 10^{10}
\end{array}$ \\
\hline

$\tilde{m}_1 (eV)$ &
$\begin{array}{r}
0.19\\
0.11
\end{array}$ &
$\begin{array}{r}
0.78\\
0.48
\end{array}$ &
$\begin{array}{r}
5.17\\
3.18
\end{array}$ &
$\begin{array}{r}
1.19\\
0.7
\end{array}$ \\
\hline

$Y_B^{max}$ &
$\begin{array}{r}
1.0 \cdot 10^{-8}\\
3.6\cdot 10^{-10}
\end{array}$ &
$\begin{array}{r}
1.2 \cdot 10^{-10}\\
4.3 \cdot 10^{-12}
\end{array}$ &
$\begin{array}{r}
2.8 \cdot 10^{-11}\\
9.7 \cdot 10^{-13}
\end{array}$ &
$\begin{array}{r}
6.6 \cdot 10^{-10}\\
2.3 \cdot 10^{-11}
\end{array}$ \\
\hline
$Y_B^{*}$ &
$\begin{array}{r}
1.3 \cdot 10^{-10}\\
2.8 \cdot 10^{-12}
\end{array}$ &
$\begin{array}{r}
3.5  \cdot 10^{-11}\\
7 \cdot 10^{-13}
\end{array}$ &
$\begin{array}{r}
1.2 \cdot 10^{-12}\\
2.6 \cdot 10^{-14}
\end{array}$ &
$\begin{array}{r}
3.2\cdot 10^{-12}\\
6.9  \cdot 10^{-14}
\end{array}$ \\
\hline\hline
\end{tabular}
\end{center}
\caption{\it Baryon asymmetry predictions based on four representative
fits. Here, $Y_B^{max}$ is the value obtained using
eq.(\ref{eq:ebound}), and $Y_B^*$ 
is the prediction for $Y_B$ using  eq. (\ref{eq:eps}) and inserting a phase of 0.1 rad in the 
(12) element of $Y_\nu$.
In each row the upper value corresponds to $M_3 = 5\cdot  10^{13}$~GeV and the lower to 
 $M_3 = 10^{12}$~GeV.} 
\label{tab:table3}
\end{table}

Looking at the predictions for leptogenesis in more detail and using
the complete expression for $\epsilon_1$ in (\ref{eq:eps}), we see that fit $(i)$ can 
accommodate comfortably the observed baryon asymmetry $Y_B$ with phases of
  $\mathcal{O}$(0.1) rad, which would not change the LFV predictions. 
The remaining three models, if the phases are small, would under-produce $Y_B$.
We also note that decreasing the scale $M_3$ would decrease both $Y_B^*$ and the LFV effects,
whereas increasing $M_3$ to values that would correspond to the perturbative limit for
$Y_\nu$ would increase $Y_B^*$, but not sufficiently to reach its  experimental value with
small phases~\footnote{Furthermore, in the case of benchmark point (a) we can see in Fig. 2 that 
BR($\mu\rightarrow e \gamma$) already sets 
the upper limit of $M_3$ below $10^{14}$~GeV.}. In these cases, either one would have to postulate an additional source
of baryon asymmetry, or one should explore predictions for LFV in the presence of large phases.
Examining these possibilities lies beyond the scope of this paper. However, we do note that
overproduction of baryons is not a problem in our scenario, even in the absence of
extra sources of entropy.

\section{Conclusions}
\label{sec:5}

Abelian flavour symmetries provide interesting possibilities for
understanding the hierarchy of fermion masses and mixing. Despite 
uncertainties in the choice of ${\cal O}(1)$ coefficients, they offer useful
insight into physical observables, and provide specific
predictions for the signals to be expected in various detection channels,
which serve as diagnostic tools for discriminating between different models; moreover, several structures predicted by
non-abelian symmetries can be well reproduced by simple abelian constructions.

In our work, we have explored these possibilities, using updated experimental input
from neutrino data, particularly recent measurements of $\theta_{13}$, MEG and the LHC.
We have revisited the signatures of charged LFV within an SU(5) GUT 
framework supplemented by an abelian flavour symmetry,
studying the correlations arising in CMSSM models with parameter values that are favoured by the LHC
and cosmological considerations, finding interesting possibilities even within this most constrained scenario.
 Because of
their sensitivity to flavour symmetries and model parameters that
are not constrained by the neutrino data, particularly those linked
to the right-handed neutrino sector, LFV searches may become a
powerful tool for distinguishing between different
theoretical scenarios. 

We first performed a scan of different fits to the neutrino data,
selecting 
representative fits that lead to normal neutrino hierarchies and
correlations between the neutrino mixing angles that are
compatible with  the global analysis of neutrino data in~\cite{neu-fits}. In doing so, we
paid attention to the naturalness of the fit, avoiding artificial cancellations arising
from specific choices of coefficients. 

We then looked at the expectations for LFV processes in the above models,
identifying the range of parameters where observable signatures are possible.
In general, fits with similar predictions 
for the neutrino parameters may lead to 
different LFV predictions. However, 
the recent input  on $\theta_{13}$, combined with the new MEG bound on
$\mu \rightarrow e \gamma$ as well as LHC data,
does constrain the allowed structures.
Further precision in the determination of 
neutrino parameters could lead to 
restrictions on the choices of model coefficients,
but would not constrain the right-handed neutrino charges.
New input in this respect, however, could be provided by the rates 
for LFV processes, since their magnitude is directly linked
to these charges, unlike the neutrino mass 
and mixing parameters. Additional input on the right-handed
neutrino sector could be obtained by requiring successful
leptogenesis, which in our case can be achieved for a natural choice
of parameters.

In the cases we studied, it was possible to establish correlations between the
expected rates for radiative LFV decays, the LFV decay of the second lightest neutralino $\chi_2$ at the LHC 
and LFV in slepton decay at a future LC, 
for different
possibilities for the structure of the heavy Majorana neutrino masses.  
 
Within the CMSSM, the absence of a
supersymmetry signal at the LHC data and the discovery of 
a neutral Higgs weighing $\sim 125$~GeV imply that
observation of slepton flavour violation at 
the LHC would be difficult but possible, for points with a 
lighter spectrum. Observation of LFV at the
 LC is also possible
for the centre of mass energies above 1 TeV that 
are compatible with the nominal energies of CLIC.
On the other hand, it should be noted here that 
scenarios less constrained than the CMSSM (which could fit $m_h$ with a lighter sparticle 
spectrum) might predict observable LFV signals at even smaller energies.

\vspace*{0.3 cm}
{\bf Acknowledgements} 
The work of J.E. is supported partly by the London
Centre for Terauniverse Studies (LCTS), using funding from the European
Research Council 
via the Advanced Investigator Grant 267352. S.L. and M.E.G thank the CERN
Theory Division for its kind hospitality and acknowledges support from
the ERC Advanced Investigator Grant 267352.
The work of M.C. is supported by MultiDark  under Grant No. CSD2009-00064 of the
Spanish MICINN  Consolider-Ingenio 2010 Program. 
M.E.G. and M.C. acknowledge further support from the
MICINN project FPA2011-23781 and  the Grant MICINN-INFN(PG21)AIC-D-2011-0724.



\end{document}